\documentstyle[prb,aps,preprint,tighten]{revtex}

\ifpreprintsty
\else
\newcommand{\startlongequation}{
\end{multicols}\vspace*{-3.5ex}{\tiny\noindent
\begin{tabular}[t]{c|} \parbox{0.493\hsize}{~} \\ \hline \end{tabular}} }

\newcommand{\stoplongequation}{
{\tiny\hspace*{\fill}
\begin{tabular}[t]{|c}\hline\parbox{0.49\hsize}{~} \\ \end{tabular}}
\vspace*{-2.5ex}\begin{multicols}{2} }

\fi

\begin{document}
\widetext

\draft

\title{Proof for an upper bound in fixed-node Monte Carlo for lattice
fermions}

\author{D.F.B. ten Haaf, H.J.M. van Bemmel, J.M.J. van Leeuwen,
and W. van Saarloos}
\address{Institute Lorentz, Leiden University,
 P. O. Box 9506, 2300 RA Leiden, The Netherlands}
\author{D.M. Ceperley}
\address{NCSA, University of Illinois at Urbana-Champaign,
405 N. Mathews Avenue, Urbana, Illinois 61801, USA}

\ifpreprintsty
\date{Received December 7, 1994}
\else
\date{Preprint INLO-PUB-21/94, December 7, 1994}
\fi

\maketitle

\begin{abstract}
  We justify a recently proposed prescription
for performing Green Function Monte Carlo calculations
on systems of lattice fermions, by which
one is able to avoid the sign problem. We generalize the prescription
such that it can also be used for problems with hopping terms of
different signs.
We prove that the
effective Hamiltonian, used in this method, leads to an
upper bound for the ground-state energy of the real
Hamiltonian, and we illustrate the effectiveness of the
method on small systems.
\end{abstract}


\ifpreprintsty
\else
\begin{multicols}{2}
\fi
\section{Motivation}
As is well known, exact Monte Carlo methods cannot be applied straightforwardly
to fermionic systems. In such systems, the sign problem causes great
difficulties in obtaining sufficient statistical accuracy,\cite{statacc}
particularly as the number of quantum particles increases. The
reason is that, when sampling physical properties in configuration
space, one collects large positive and negative contributions,
due to the fact that a fermion wave function is of different
sign in different regions of the configuration space. These
contributions tend to cancel, giving a result that may be
exponentially smaller than the positive and negative contributions
separately.

Recently, some of us described a method to perform
Green Function Monte Carlo (GFMC) on a system of fermions
on a lattice\cite{bemmel}, which is an extension of the
fixed-node Monte Carlo method for continuum problems,
developed by Ceperley and Alder\cite{ceperley}.
In this method
one avoids the sign problem, replacing the original
Hamiltonian by an effective
Hamiltonian,
such that one obtains contributions of one sign
only in the sampling procedure.
The price one has to pay is in the fact that the ground-state
energy $E_{\text{eff}}$
of the effective Hamiltonian is in general
not the same as the
ground-state energy $E_{\text{0}}$
of the original Hamiltonian.
It was claimed, however, that $E_{\text{eff}}$ is a true upper
bound for $E_{\text{0}}$, making the method variational.

The proof for this
upper bound is less obvious than was suggested in Ref.\onlinecite{bemmel},
because an assumption was used about the form of the ground state of the
effective Hamiltonian, which is not generally true [see the discussion
following Eq.~(\ref{leverrule})]. It is possible,
however, to give a general proof for the upper bound.
In the process of deriving this proof, we found that our
prescription can be generalized, such that also problems
with a Hamiltonian containing hopping terms of different signs
can be treated by this method.
The aim of this paper is to give a general proof, illustrate the
method on
small systems, for which we diagonalize both the original and the
effective Hamiltonians exactly, and discuss the
applicability of the method. We do not actually perform
Monte Carlo simulations here.

\section{Effective Hamiltonian}
  We work in a configuration space $\{R\}$, where each $R$ denotes
a configuration of numbered fermions on a lattice. In this configuration
space, the Hamiltonian ${\cal H}$ of our problem
can be represented by a real
symmetric matrix $H$ with elements
$\langle R|H|R^\prime\rangle$. One is generally interested in finding the
ground-state energy of this Hamiltonian subject to some symmetry constraints,
for example that the wavefunction be antisymmetric.
We suppose that the
ground state $|\psi_{\text{0}}\rangle$ of ${\cal H}$ is reasonably
well approximated by a trial state $|\psi_{\text{T}}\rangle$,
which is defined through its wave function in all possible
configurations: $\psi_{\text{T}}(R) = \langle R|\psi_{\text{T}}\rangle$.
We restrict ourselves to real trial wave functions, because the ground-state
wave function
can be taken real in this problem, and the sign of the trial wave
function is one of the key ingredients for our method.
Complex Hamiltonians and trial functions can be treated with the
so-called {\em fixed-phase} method\cite{fixedphase}.
Typical examples of the Hamiltonians considered here are the Hubbard
Hamiltonian or
the Kondo lattice model, and the typical trial wave function is
a determinant obtained by a mean-field approximation.

  In general, a trial wave function divides the configuration
space into {\em nodal regions\/}. A nodal region is a set of
configurations in which the trial wave function has the same
sign, and which are connected via the Hamiltonian. For an
antisymmetric wave function, there is equivalence between
the regions of positive and the regions of negative sign.

  In GFMC with {\em importance sampling},
random walkers diffuse and branch through the configuration space
in a stochastic way, guided by a trial wavefunction.
The Hamiltonian is used to project out the lowest energy state.
The process for a lattice model and its mathematical justification
is described in more
detail in Appendix~A and Ref.\onlinecite{tricep}.
In the fixed-node approach, one ensures that the contribution
of a specific walker is always positive, otherwise the
negative signs will eventually interfere destructively. For completeness, and
to indicate the connection with the sign problem in quantum
Monte Carlo simulations, we expand on
this point in Appendix~A. If  the off-diagonal terms  in the Hamiltonian
are all negative (as in the Hubbard model), a sign change
only occurs when a walker goes from one nodal region to the other.
More generally, a walker could collect an unwanted minus sign if there
exists a pair of configurations $R$ and $R^\prime$ such that:
\begin{equation}\label{nonhoppingcondition}
 \langle R|H|R^\prime\rangle\psi_{\text{T}}(R)\psi_{\text{T}}(R^\prime) > 0 .
\end{equation}
In order to prevent this from happening, we make an effective Hamiltonian
which does not have such matrix elements.

  The fixed-node method was developed for the case in which the electron
coordinates are continuous variables\cite{anderson,ceperley}.
There, one has to deal with kinetic terms of negative sign only, and
the nodal surface of a trial wave function is
uniquely defined as the set of configurations where it vanishes.
The fixed-node constraint can be implemented
by imposing the boundary condition that $\psi$ must vanish on the
nodal surface of $\psi_{\text{T}}$. In the limit of sufficiently
small step sizes, we can make sure that Eq.~(\ref{nonhoppingcondition}) is
never
violated since $R$ and $R^\prime$ become closer together and $\psi_{\text{T}}$
will vanish.
In this way one obtains the lowest energy under the condition
that the wave function has the same nodal structure
as the trial wave function. This energy yields an upper bound to
the true ground-state energy; in practice, very accurate estimates
for the ground-state energy of continuum problems
can be obtained.

On a lattice,
one has to deal with discrete steps, and one has to treat the hops
that cause a change of sign in a different way.
In our implementation, we replace those unwanted hopping
terms in the Hamiltonian by diagonal terms, that depend on the
ratio of the trial wave
function in the configurations $R$ and $R^\prime$. We
thus construct an effective Hamiltonian ${\cal H_{\text{eff}}}$ as follows:
\begin{eqnarray}
 \langle R|H_{\text{eff}}|R^\prime\rangle & = &
 \langle R|H|R^\prime\rangle \hspace*{0.5em}
 \mbox{(if $\langle
R|H|R^\prime\rangle\psi_{\text{T}}(R)\psi_{\text{T}}(R^\prime) < 0$)} \nonumber
\\
  & = &
 0 \hspace*{4.1em} \mbox{(otherwise)} \label{Heffoffdiag}
\end{eqnarray}
are the off-diagonal terms, and the diagonal terms are given by
\begin{equation}\label{Heffdiag}
 \langle R|H_{\text{eff}}|R\rangle = \langle R|H|R\rangle +
  \langle R|V_{\text{sf}}|R\rangle .
\end{equation}
The last term in~(\ref{Heffdiag}) is the {\em sign-flip
potential\/} at $R$, which
corrects for the contributions of the steps
left out in $H_{\text{eff}}$. $V_{\text{sf}}$ has only diagonal elements,
which are defined by
\begin{equation}\label{Vsf}
\langle R|V_{\text{sf}}|R\rangle = \sum_{R^\prime}^{\text{sf}}\langle
R|H|R^\prime\rangle\frac{\psi_{\text{T}}(R^\prime)}{\psi_{\text{T}}(R)} .
\end{equation}
Here the summation is over all [neighboring] configurations $R^\prime$
of $R$ for which~(\ref{nonhoppingcondition}) holds.
Note that this is a significant extension of the prescription presented
in Ref.\onlinecite{bemmel}, where we only considered the case that all
hopping terms are of negative sign, such that the sign-flipping hops
would coincide exactly with the hops to a different nodal region.
In the general case, we prefer to speak about sign-flipping steps
instead of nodal-boundary steps, as the latter term may cause confusion.

  Clearly, by this prescription, a hop that would induce a sign change
is replaced
by a positive diagonal potential.
If instead one used only the truncated Hamiltonian
as given by~(\ref{Heffoffdiag}), with the original diagonal matrix elements
$\langle R|H_{\text{eff}}|R\rangle = \langle R|H|R\rangle$, then
the value of the wave function at the
node would be too high and its energy too low.
This was found in an
earlier attempt to perform fixed-node Monte Carlo on
lattice fermions by An and van Leeuwen.\cite{an}

  A somewhat similar procedure, called ``model-locality'',
has been used by Mitas {\em et al.}\cite{mitas} in continuum problems with a
non-local potential
that arises from replacing atomic cores with pseudopotentials.
As in a lattice system, they can not solve the problem of crossing a node
by making the step size of the walkers continuously smaller, because of
the non-local potential that connects configurations at finite distances.
In their approach the unwanted off-diagonal terms are truncated, and
replaced by diagonal contributions as in Eq.~(\ref{Vsf}), but with the sum
over all $R^\prime$, not just over sign-flip configurations.
With the model-locality procedure,
one does not obtain an upper bound for the ground-state energy.

\section{Upper bound}
We want to show that the prescription given above for ${\cal H}_{\text{eff}}$
leads to an upper bound for the ground-state
energy of ${\cal H}$. In order to do so, we define a truncated Hamiltonian
${\cal H}_{\text{tr}}$, and a {\em sign-flip\/} Hamiltonian ${\cal
H}_{\text{sf}}$, by
\begin{eqnarray}
 {\cal H} & = & {\cal H}_{\text{tr}} + {\cal H}_{\text{sf}} ,\\
 {\cal H}_{\text{eff}} & = & {\cal H}_{\text{tr}} + {\cal V}_{\text{sf}} ,
\end{eqnarray}
where the diagonal elements of $H_{\text{tr}}$ are
\begin{equation} \langle R|H_{\text{tr}}|R\rangle = \langle R|H|R\rangle ,
\end{equation}
and its off-diagonal elements are given by
\begin{equation} \langle R|H_{\text{tr}}|R^\prime\rangle = \langle
R|H_{\text{eff}}|R^\prime\rangle .\end{equation}
${\cal V}_{\text{sf}}$ is the sign-flip potential, for which the
matrix elements are given by~(\ref{Vsf}), and $H_{\text{sf}}$ contains only the
off-diagonal elements of $H$ which are put to zero in the effective
Hamiltonian.
We now take {\em any\/} state
\begin{equation} |\psi\rangle = \sum_R |R\rangle\psi(R) ,
\end{equation}
 and we compare its energy with respect to ${\cal H}$ and to ${\cal
H}_{\text{eff}}$:
\begin{eqnarray} \nonumber
 \Delta E & = & \langle\psi|({\cal H}_{\text{eff}}-{\cal H})|\psi\rangle \\
          & = & \langle\psi|({\cal V}_{\text{sf}}-{\cal
H}_{\text{sf}})|\psi\rangle .
\end{eqnarray}
$\Delta E$ can be written explicitly in terms of the matrix elements of
$V_{\text{sf}}$ and $H_{\text{sf}}$:
\ifpreprintsty
\else
\startlongequation
\fi
\begin{equation}
 \Delta E = \sum_R\psi(R)^\ast\left[
     \langle R|V_{\text{sf}}|R\rangle\psi(R)
   - \sum_{R^\prime}\langle R|H_{\text{sf}}|R^\prime\rangle\psi(R^\prime)
                           \right] .
\end{equation}
We rewrite this expression in terms of the matrix elements of $H$:
\begin{equation}\label{doublesum}
 \Delta E = \sum_R\psi(R)^\ast\left[
     \sum_{R^\prime}^{\text{sf}}\langle
R|H|R^\prime\rangle\frac{\psi_{\text{T}}(R^\prime)}{\psi_{\text{T}}(R)}\psi(R)
   - \sum_{R^\prime}^{\text{sf}}\langle R|H|R^\prime\rangle\psi(R^\prime)
                           \right] .
\end{equation}
In this double summation each pair of configurations $R$ and $R^\prime$ occurs
twice. We combine
these terms and rewrite~(\ref{doublesum}) as a summation over pairs:
\begin{equation}
\Delta E = \sum_{(R,R^\prime)}^{\text{sf}}\langle R|H|R^\prime\rangle\left[
\left|\psi(R)\right|^2\frac{\psi_{\text{T}}(R^\prime)}{\psi_{\text{T}}(R)} +
\left|\psi(R^\prime)\right|^2\frac{\psi_{\text{T}}(R)}
{\psi_{\text{T}}(R^\prime)} - \psi(R)^\ast\psi(R^\prime) -
\psi(R^\prime)^\ast\psi(R)
                           \right] .
\end{equation}
Denoting by $s(R,R^\prime)$ the sign of the matrix element $\langle
R|H|R^\prime\rangle$, and using the fact that for all terms in this
summation the condition~(\ref{nonhoppingcondition}) is satisfied, we can
finally write $\Delta E$ as
\begin{equation} \label{deltaEfinal}
\Delta E = \sum_{(R,R^\prime)}^{\text{sf}}\left|\langle
R|H|R^\prime\rangle\right|
\left|\psi(R)\sqrt{\left|\frac{\psi_{\text{T}}(R^\prime)}
{\psi_{\text{T}}(R)}\right|} -
s(R,R^\prime)\psi(R^\prime)\sqrt{\left|\frac{\psi_{\text{T}}(R)}
{\psi_{\text{T}}(R^\prime)}\right|}\right|^2 .\end{equation}
\ifpreprintsty
\else
\stoplongequation\noindent
\fi
Note that we do not have to worry about configurations $R$
where $\psi_{\text{T}}(R)=0$: they do not occur in this summation.
Obviously, $\Delta E$ is positive for any wave function $\psi$. Thus the
ground-state energy of ${\cal H}_{\text{eff}}$ is an upper bound for the
ground-state energy of the original Hamiltonian ${\cal H}$.

Now the GFMC method can calculate the exact ground-state energy
$E_{\text{eff}}$ and
wavefunction $\psi_{\text{eff}}$
of $H_{\text{eff}}$, without any sign problem.
Assuming the trial function  $\psi_{\text{T}}$ has the correct symmetry
[for example is antisymmetric], then
$\psi_{\text{eff}}$ will carry the same symmetry and hence:
$E_{\text{eff}} \ge \langle\psi_{\text{eff}} | H | \psi_{\text{eff}}\rangle \ge
E_0$,
where the second inequality follows from the usual variational principle.
Hence the fixed-node energy is an upper bound to the true ground-state energy.
One can easily verify that ${\cal H}|\psi_{\text{T}}\rangle = {\cal
H}_{\text{eff}}|\psi_{\text{T}}\rangle$, and thus one can be sure that
the GFMC procedure improves on the energy of the trial wave function:
$E_{\text{eff}} \le \langle\psi_{\text{T}} | H_{\text{eff}}
|\psi_{\text{T}}\rangle = \langle\psi_{\text{T}} | H | \psi_{\text{T}}\rangle$.

\section{Variation of the trial state}
Let us consider the situation where we use the exact ground state
$|\psi_{\text{0}}\rangle$ of ${\cal H}$, with energy $E_0$, as trial state.
Obviously, for the method to be useful, it is desirable
that in that case the effective Hamiltonian has the same ground-state
energy $E_0$, and the same ground state $|\psi_{\text{0}}\rangle$, as that
would
make it possible to find the true ground state by varying the
trial wave function in some way. In Eq.~(\ref{deltaEfinal})
we substitute $\psi_{\text{0}}$ for $\psi_{\text{T}}$.
In order to have $\Delta E$ equal to zero, each individual term
in the summation~(\ref{deltaEfinal}) has to vanish, thus leading to
\begin{equation}
 \psi(R)\sqrt{\left|\frac{\psi_{\text{0}}(R^\prime)}
 {\psi_{\text{0}}(R)}\right|} -
s(R,R^\prime)\psi(R^\prime)\sqrt{\left|\frac{\psi_{\text{0}}(R)}
{\psi_{\text{0}}(R^\prime)}\right|} = 0 ,
 \end{equation}
 or,
\begin{equation}
 \frac{\psi(R)}{\psi(R^\prime)} =
s(R,R^\prime)\left|\frac{\psi_{\text{0}}(R)}{\psi_{\text{0}}(R^\prime)}\right|
  = \frac{\psi_{\text{0}}(R)}{\psi_{\text{0}}(R^\prime)}
 \end{equation}
for all sign-flipping pairs $(R,R^\prime)$.
This condition is trivially fulfilled for $\psi=\psi_{\text{0}}$. Thus, the
true ground-state energy can be reached by variation
of the trial wave function.
One can further extend
this result to show that as $\psi_{\text{T}} \rightarrow \psi_{\text{0}}$ the
error in the fixed-node
energy will be second order in the difference, $\psi_{\text{T}} -
\psi_{\text{0}}$, with the
coefficient  positive.

  The original contention in Ref.\onlinecite{bemmel} was that the wave
function,
obtained through this effective Hamiltonian, would have exactly the same
ratio at each sign-flipping pair $R$ and $R^\prime$
as the trial wave function, i.e.
\begin{equation}\label{leverrule}
  \frac{\psi_{\text{eff}}(R)}{\psi_{\text{eff}}(R^\prime)} =
\frac{\psi_{\text{T}}(R)}{\psi_{\text{T}}(R^\prime)} .
\end{equation}
This, however, is in general not the
case, as one can see from considerations about the symmetry of
the wave function.
An example for a
small system, which illustrates this point, is given in Appendix~B.
Note that our proof for the upper bound does not rely on the
assumption~(\ref{leverrule}),
and that the conclusion we put forward in Ref.\onlinecite{bemmel} about
the variational principle, remains unchanged. In fact, because the
ground state of ${\cal H}_{\text{eff}}$ is found in a much less
restricted space of states than those satisfying~(\ref{leverrule}),
the resulting estimate for the ground-state energy is much better
than was anticipated.

There is an important difference between the lattice and continuum fixed-node
method.  In the continuum method, it is only the sign of the trial function
that matters.  If the nodes are correctly placed, one will obtain the
exact energy regardless of the magnitude of the trial function.  Clearly
this does not hold with the lattice fixed-node procedure: the sign of the
trial function
and the relative magnitudes of the trial function in
configurations that are connected by a sign flip
must be correct.
For example, in the continuum the exact
result would be obtained for a one-dimensional problem since the
nodal surface are the coincident hyperplanes.
One does not nessecarily get the exact result for a 1-d lattice model as one of
the following examples shows.

\section{Illustrations}
We illustrate the effect of the effective Hamiltonian for the
single-band Hubbard model by
exact calculations on
two small systems: a loop of 4 lattice sites on the corners
of a square, and a graph consisting of 8 points on the corners of a cube.
We use the well-known Hamiltonian
\begin{equation}
 {\cal H} = -t\sum_{\langle i,j\rangle ,\sigma}
c_{i\sigma}^{\dagger\rule[-.5ex]{0cm}{2ex}}
c_{j\sigma}^{\rule[-.5ex]{0cm}{2ex}} + U\sum_i
n_{i\uparrow}^{\rule[-.5ex]{0cm}{2ex}}n_{i\downarrow}^{\rule[-.5ex]{0cm}{2ex}}
,
\end{equation}
containing nearest-neighbor hopping with strength $t$ and a local interaction
between electrons of opposite spin with strength $U$. We consider the square
with 2 electrons with spin up and 2 with spin down, and the cube with 4 up and
1 down and with 4 up and 4 down, respectively. In all cases, we use the results
of self-consistent mean-field calculations as trial wave functions, and we
compare the mean-field energy (MF), the lowest energy of the effective
Hamiltonian
(FN, for fixed node), and the exact ground-state energy, for different values
of the interaction parameter $U$. We use different restrictions on the average
number
of electrons with spin up and down per site, in order to obtain different
types of self-consistent mean-field wave functions. Writing
$\langle{n_{i\sigma}}\rangle = \langle{n_{\sigma}}\rangle +
(-1)^{\sigma}q_{i\sigma}$ for the average number of spin-$\sigma$ particles on
site $i$, we denote $q_{i\sigma} = 0$ by H (homogeneous), and $q_{i\sigma} =
(-1)^iq_{\sigma}$, with $q_{\sigma}$ a constant, by AF (antiferromagnetic, or
N\'eel order favoured). In the case of the cube with 4 up and 1 down spins
[i.e. off half filling], the self-consistent mean-field solution with lowest
energy
turns out to have a symmetry different from both H and AF. As the exact ground
state
of ${\cal H}$ as well as of ${\cal H}_{\text{eff}}$ is not degenerate in these
cases, it cannot have broken symmetry for $\langle{n_{i\sigma}}\rangle$.
Note that, for $U=0$, the mean-field approximation yields the exact ground
state, and we checked that
also the fixed-node result equals the exact ground-state energy in that case.
The results are presented in Table~\ref{tab-results}.

As one can see, the fixed-node approach on these small systems yields a
significant improvement on the upper bound for the ground-state energy,
compared to the mean-field approximations. One may note
the fact that the mean-field wave function with lowest energy
does in general not give
the best fixed-node result. In a real problem, one would want to
find the best possible trial wave function as input for the fixed-node
procedure, and it is clear from these results that `best' does {\em not\/}
mean `having the lowest variational
energy' here. The sign of
the trial wave function and its behavior at the nodal boundary determine
how good the fixed-node energy will be.

\ifpreprintsty
\else
\begin{figure}
\refstepcounter{table}
\label{tab-results}
\begin{center}\vspace*{-2ex}
\begin{tabular}{|c|c|c|r@{}l|r@{}l|r@{}l|}
\hline
 system & $U$ & trial & \multicolumn{6}{c|}{energies} \\
 \cline{4-9}
 & & type & \multicolumn{2}{c|}{MF} & \multicolumn{2}{c|}{FN} &
\multicolumn{2}{c|}{exact} \\
 \hline \hline
 square & $0$ & & -4 & & -4 & & -4 & \\
 2$\uparrow$~~2$\downarrow$ & $1$ & AF & -3 & .2855 & -3 & .3172 & -3 & .3409
\\
 \hline
 cube & $0$ & & -9 & & -9 & & -9 & \\
 4$\uparrow$~~1$\downarrow$ & $1$ & H & -8 & .5 & -8 & .5419 & -8 & .5420 \\
  & $6$ & H & -6 & & -7 & .2508 & -7 & .2533 \\
  & $6$ &  & -6 & .0701 & -7 & .2424 & -7 & .2533 \\
  & $10$ & H & -4 & & -6 & .8400 & -6 & .8442 \\
  & $10$ & AF & -4 & .2551 & -6 & .7476 & -6 & .8442 \\
  & $10$ &  & -5 & .3271 & -6 & .7637 & -6 & .8442 \\
 \hline
 cube & $0$ & & ~-12 & & ~-12 & & ~-12 & \\
 4$\uparrow$~~4$\downarrow$ & $1$ & H & -10 & & -10 & .1148~ & -10 & .1188~ \\
  & $~2.5~$ & H & -7 & & -7 & .7257 & -7 & .7510 \\
  & $2.5$ & AF & -7 & .0061~ & -7 & .6942 & -7 & .7510 \\
  & $10$ & H & 8 & & -2 & .6597 & -2 & .8652 \\
  & $10$ & AF & -2 & .3113 & -2 & .6382 & -2 & .8652 \\
 \hline
\end{tabular} \\[2mm]
\parbox{\hsize}{TABLE~\ref{tab-results}. Comparison of the energies obtained
for three different systems by means of self-consistent mean-field (MF),
fixed-node (FN), and exact calculations\cite{exact}. All values are given in
units of $t$. }
\end{center}\vspace*{-1ex}
\end{figure}
\fi

\section{Conclusions and outlook}
This method can be applied to any lattice model
provided that only the Hamiltonian and a trial wave function with
the proper symmetry are given.
It is super-variational, in the sense that it always yields an upper bound
for the energy,
which is the lowest possible value consistent with the imposed constraint.
By varying the the effective Hamiltonian through the
trial wave function, in principle the exact ground-state energy can
be obtained.

Note that we have not used the symmetries of the trial state as input
for our method. This means that it is possible to use this method for
models of frustrated spins on a lattice and, via the appropriate mappings,
for systems of bosons as well, or for excited states which are ground
states of a given symmetry. In a forthcoming publication,
possibilities to do so will be presented and discussed.

Note further that the {\em nodal relaxation\/} method, as described in
Ref.\onlinecite{ceperley} for continuum problems, is also applicable on
the lattice. In this method, one uses the fixed-node approach to improve
on the trial wave function. When this has been done, one removes the
sign-flip constraint, and allows the walkers to move through the whole
configuration space. If the fixed-node result is close enough to
the ground state, one can sample the exact ground-state
energy before the sign problem destroys the accuracy.

In the near future, we plan to use this method for Monte Carlo studies on some
of the systems mentioned above, in order to find more comparisons of the
fixed-node approach with known results, to check the effectiveness of the
method, and to tackle some new problems as well.

The method appears to be also useful for continuum problems, where one
has a non-local potential. For example one can modify the model-locality
approach of Ref.\onlinecite{mitas} so that it does yield an upper bound.
Essentially one must allow non-local moves which do not change the sign and
add terms to the effective Hamiltonian corresponding to discarded moves.

We finally note that another promising avenue for further development of the
conceptual basis of our approach is given by the observation by Martin that the
use of an effective Hamiltonian can be
couched in the language of Density Functional Theory.\cite{martin} This makes
it possible to apply a number of well-known results for the behavior
of the energy functional under variation of both the effective Hamiltonian
and the trial state $|\psi_{\text{T}}\rangle$.

\section*{acknowledgements}
The authors want to thank
P.J.H. Denteneer for stimulating discussions. This research was supported by
the Stichting Fundamenteel Onderzoek der Materie (FOM), which is financially
supported by the Nederlandse Organisatie voor Wetenschappelijk Onderzoek (NWO).
DMC was supported by the Institute for Theoretical Physics at the University
of California at Santa Barbara.

\section*{Appendix A: \ifpreprintsty \else \\ \fi The sign problem in Monte
Carlo}
\setcounter{equation}{0}
\renewcommand{\theequation}{A\arabic{equation}}

In this Appendix, we clarify the origin of the sign problem for a
specific way of performing Green Function Monte Carlo on lattice
fermions, and we explain how one
is able to circumvent this problem, using the fixed-node approach.
More details of this version of GFMC as applied to lattices are
given in Ref.\onlinecite{tricep}.

In a GFMC simulation one tries to obtain
information about the properties of the ground state of a given
Hamiltonian ${\cal H}$. Starting from a trial state, one can
obtain [a stochastic representation of] the ground state by
repeatedly applying a projection [or diffusion] operator.
On a lattice it is simplest to use
an operator that is linear in ${\cal H}$, and
that can be viewed as the first-order expansion of an exponential
diffusion operator in imaginary time:
\begin{equation}
{\cal F} = 1-\tau({\cal H} - w) ,
\end{equation}
where $w$ is a parameter that should be chosen close to the
ground-state energy in order to keep the wavefunction normalized.
The parameter $\tau$ is taken small enough to ensure
that the diagonal terms
of this operator are positive. The off-diagonal elements in the
matrix representation for ${\cal F}$ are, up to a factor $-\tau$, the
same as those for ${\cal H}$.
The $n$-th approximation of the ground state
is given by
\begin{equation}
|\psi^n\rangle = {\cal F}^n|\psi_{\text{T}}\rangle .
\end{equation}
One can check that, if the
trial state has some overlap with the ground state, $|\psi^n\rangle$
will converge exponentially fast to the ground state for large $n$.

The ground-state energy can be calculated as:
\begin{equation}
 E_n = \frac{\langle\psi_{\text{T}}|{\cal
H}|\psi^n\rangle}{\langle\psi_{\text{T}}|\psi^n\rangle}
\end{equation}
We rewrite this expression
as a summation over paths in configuration space:
\begin{equation}
 E = \frac{\rule[-1.0ex]{0cm}{2ex}\sum_{\cal
R}E(R_n)\langle\psi_{\text{T}}|R_n\rangle\prod_{i=1}^{n}
 \langle R_i|F|R_{i-1}\rangle \langle R_0|\psi_{\text{T}}\rangle }
          {\rule[.1ex]{0cm}{2ex}\sum_{\cal
R}\langle\psi_{\text{T}}|R_n\rangle\prod_{i=1}^{n}
 \langle R_i|F|R_{i-1}\rangle \langle R_0|\psi_{\text{T}}\rangle } ,
\end{equation}
where
$E(R)\equiv\langle\psi_{\text{T}}|H|R\rangle
\langle\psi_{\text{T}}|R\rangle^{-1}$
is the local energy at $R$, and
${\cal R} = \{R_{0},R_{1},R_{2},...,R_{n}\}$
denotes a path in configuration space.

In a GFMC procedure this expression is sampled stochastically
by constructing paths ${\cal R}$ in configuration
space, and calculating the energy from the contributions of those paths.
Importance sampling is used to reduce the fluctuation of those paths
by modifying ${\mathcal F}$.
The sign problem arises from the fact since the fermion trial wave function
is antisymmetric, its sign will vary.
Also, the matrix elements $\langle R|F|R^\prime\rangle$ between different $R$
and $R^\prime$ need not be
always positive. Thus, when performing a random walk to obtain a
path ${\cal R}$, starting from a configuration $R_0$ where the trial
wave function $\psi_{\text{T}}(R_0)$ is of specific sign, one may
end up in a configuration $R_n$ where the trial function is of
the opposite sign, or one may have collected an odd number of
negative $\langle R|F|R^\prime\rangle$ in
the path. For large $n$, one obtains about as many positive
as negative contributions; the difference is used to determine
the energy. One can easily show that the ``signal-to-noise'' ratio
must decrease exponentially in $n$ once negative contributions are allowed.
Intuitively, it is easy to understand that this will
give rise to an inaccurate result. In practice, this severely
limits the applicability of Quantum Monte Carlo methods to fermion
problems.

In the fixed-node approach, one wants to avoid that contributions
of different sign can be obtained. In order to ensure this, one
demands that {\em at every individual step\/} along a path,
only positive contributions are allowed.
Thus, all steps
satisfying Eq.~(\ref{nonhoppingcondition}) have to be discarded.
The prescription~(\ref{Heffoffdiag}--\ref{Vsf}) for the effective
Hamiltonian takes care of this constraint.
Finally, we remark that this prescription fits very well with the
way we perform importance sampling. When using the trial wave
function as a guiding function for the random walks, at any
point in the walk one needs to know the value of the trial
wave function, and one can use this value at the same time for
guiding the walks and for the implementation of the fixed-node effective
Hamiltonian.
Note that the summation needed to define the effective potential
in Eq.~(\ref{Vsf}) only grows linearly with the
size of the system for a Hamiltonian such as the Hubbard model.
Thus it does not appreciably slow the calculation.

\section*{Appendix B: \ifpreprintsty \else \\ \fi Example of Fixed-Node
procedure}
\setcounter{equation}{0}
\renewcommand{\theequation}{B\arabic{equation}}

In this Appendix, we give an illustration of how the effective
Hamiltonian is created, and what its effect is, on a very simple
small system. All steps can be straightforwardly generalized to
more complicated systems.

Consider the Hamiltonian
\begin{equation}
 {\cal H} = -t\sum_{\langle i,j\rangle ,\sigma}
c_{i\sigma}^{\dagger\rule[-.5ex]{0cm}{2ex}}
c_{j\sigma}^{\rule[-.5ex]{0cm}{2ex}}
\end{equation}
 on a loop of 4 sites with 2 spinless fermions. We define
{\em configurations of labeled fermions\/} $[i_1i_2]$,
where particle
$j$ ($1 \leq j \leq 2$) sits on site $i_j$ ($1 \leq i_j \leq 4$).
We number the sites, as follows:\\
\setlength{\unitlength}{0.6mm}
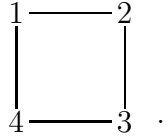
\begin{picture}(30,46)(-56,-8)
\put(6,27){\line(1,0){18}}
\put(6,3){\line(1,0){18}}
\put(3,6){\line(0,1){18}}
\put(27,6){\line(0,1){18}}
\put(3,27){\makebox(0,0){1}}
\put(27,27){\makebox(0,0){2}}
\put(27,3){\makebox(0,0){3}}
\put(3,3){\makebox(0,0){4}}
\put(35,3){\makebox(0,0){.}}
\end{picture} \\
A valid [i.e. antisymmetric] fermion wave function $\psi$ must satisfy
$\psi\left(\rule{0cm}{2ex}[ij]\right)
= -\psi\left(\rule{0cm}{2ex}[ji]\right)$.
The configuration space of this system consists of
12 configurations, and
can be depicted as follows:\\
\setlength{\unitlength}{0.8mm}
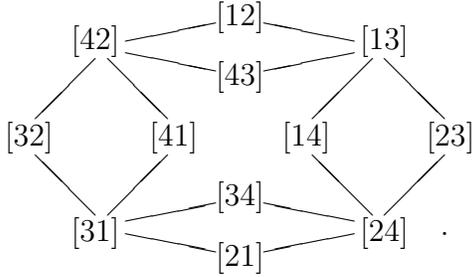
\begin{picture}(100,60)(10,0)
\thinlines
\put(42,14){\line(5,-1){15}}  % (31) to (21)
\put(42,16){\line(5,1){15}}   % (31) to (34)
\put(42,44){\line(5,-1){15}}  % (42) to (43)
\put(42,46){\line(5,1){15}}  % (42) to (12)
\put(80,46){\line(-5,1){15}}  % (13) to (12)
\put(80,44){\line(-5,-1){15}}  % (13) to (43)
\put(80,16){\line(-5,1){15}}  % (24) to (34)
\put(80,14){\line(-5,-1){15}}  % (24) to (21)
\put(37,43){\line(-1,-1){10}}  % (42) to (32)
\put(39,43){\line(1,-1){10}}  % (42) to (41)
\put(83,43){\line(-1,-1){10}}  % (13) to (14)
\put(85,43){\line(1,-1){10}}  % (13) to (23)
\put(85,17){\line(1,1){10}}  % (24) to (23)
\put(83,17){\line(-1,1){10}}  % (24) to (14)
\put(39,17){\line(1,1){10}}  % (31) to (41)
\put(37,17){\line(-1,1){10}}  % (31) to (32)
\put(37,14){\makebox(0,0){[31]}}
\put(37,46){\makebox(0,0){[42]}}
\put(85,46){\makebox(0,0){[13]}}
\put(85,14){\makebox(0,0){[24]}}
\put(61,50){\makebox(0,0){[12]}}
\put(61,40){\makebox(0,0){[43]}}
\put(61,20){\makebox(0,0){[34]}}
\put(61,10){\makebox(0,0){[21]}}
\put(26,30){\makebox(0,0){[32]}}
\put(50,30){\makebox(0,0){[41]}}
\put(72,30){\makebox(0,0){[14]}}
\put(96,30){\makebox(0,0){[23]}}
\put(95,14){\makebox(0,0){.}}
\end{picture} \\
The lines [or bonds] represent valid hops in this space. The
matrix elements $\langle [ij]|H|[kl]\rangle$ of the Hamiltonian for this system
are $-t$ if there is a bond between $[ij]$ and $[kl]$ [in that case $i=k$ or
$j=l$ must hold], or $0$ otherwise.
The ground state of this Hamiltonian
is symmetric under exchange of the particles, and we have to restrict the wave
function explicitly to be antisymmetric in order to find a valid fermion wave
function. To obtain a Hamiltonian $\underline H$ which describes the
fermion problem only, we define {\em antisymmetric states\/} $[\underline{i
j}]$, which are
antisymmetrized combinations of the configurations:
\begin{equation}
[\underline{ij}] = \frac{1}{\sqrt{2}}\left([ij] - [ji]\right) .
\end{equation}
In this way each pair of configurations $[ij]$ and $[ji]$ produces two states,
$[\underline{ij}]$ and $[\underline{ji}]$, which only
differ by their sign. One has the freedom to choose one of these states to
obtain only one state per pair of configurations, and one can calculate the
resulting
Hamiltonian for the $[\underline{ij}]$:
\begin{eqnarray}
\left\langle [\underline{ij}]|\underline H|[\underline{kl}]\right\rangle & = &
\frac{1}{2}
\sum_{\Pi_1}\sum_{\Pi_2}\mbox{sg}(\Pi_1)\mbox{sg}(\Pi_2)\left\langle
\Pi_1[ij]|H|\Pi_2[kl]\right\rangle \nonumber \\
 & = & \mbox{sg}(\Pi)\langle [ij]|H|\Pi[kl]\rangle ,
\end{eqnarray}
where $\Pi_1$ and $\Pi_2$ denote permutations of the two particles, sg gives
the sign of a permutation, and $\Pi[k l]$ is the permutation of $[kl]$ that can
be reached by one hop from $[ij]$, such that $\langle [ij]|H|\Pi[kl]\rangle =
-t$. We can again denote the Hamiltonian in a picture, representing matrix
elements $-t$ by thin lines and $+t$ by thick lines [we choose the
$[\underline{ij}]$ with $i<j$; other choices give different pictures but the
same results]:\\
\setlength{\unitlength}{0.8mm}
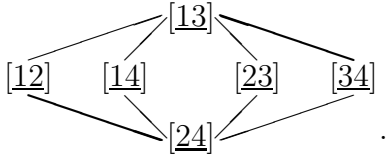
\begin{picture}(72,35)(-10,0)
\thinlines
\put(16,18){\line(3,1){22.14}}  % [12] to [13]
\put(32,18){\line(1,1){7}}  % [14] to [13]
\put(54,18){\line(-1,1){7}}  % [23] to [13]
\put(32,12){\line(1,-1){7}}  % [14] to [24]
\put(54,12){\line(-1,-1){7}}  % [23] to [24]
\put(70,12){\line(-3,-1){22.14}}  % [34] to [24]
\thicklines
\put(70,18){\line(-3,1){22.14}}  % [34] to [13]
\put(16,12){\line(3,-1){22.14}}  % [12] to [24]
\put(16,15){\makebox(0,0){[\underline{12}]}}
\put(32,15){\makebox(0,0){[\underline{14}]}}
\put(54,15){\makebox(0,0){[\underline{23}]}}
\put(70,15){\makebox(0,0){[\underline{34}]}}
\put(43,25){\makebox(0,0){[\underline{13}]}}
\put(43,5){\makebox(0,0){[\underline{24}]}}
\put(75,5){\makebox(0,0){.}}
\end{picture}\\
This structure fully contains the antisymmetry, and the corresponding
Hamiltonian gives all information there is on the fermion problem. Its
ground-state is degenerate, with energy $-2t$, and possible ground states are
 \begin{equation} |\psi_0\rangle = \frac{1}{2}\left|[\underline{13}] +
[\underline{14}] + [\underline{23}] + [\underline{24}]\right\rangle ,
\end{equation} and
 \begin{equation} |\psi_0^\prime\rangle = \frac{1}{2}\left|[\underline{12}] +
[\underline{13}] - [\underline{24}] - [\underline{34}]\right\rangle
.\end{equation}
It is easy to generalize this procedure for any system of lattice
fermions.

Let us now consider a trial state, and calculate the
effective Hamiltonian according to our fixed-node prescription.
The trial wave function defines the {\em nodal regions\/}
through its sign in all states, and, because we are working with
negative hopping terms, the sign-flip constraint reduces to
sign changes of the wave function only.
We take a very simple trial state:
\begin{equation}
 |\psi_{\mbox{\scriptsize T}}\rangle = \frac{1}{\sqrt{6}}\left|[\underline{12}]
+ [\underline{13}] + [\underline{14}] + [\underline{23}] + [\underline{24}] +
[\underline{34}]\right\rangle ,
\end{equation}
purposely chosen such that we only have to slightly adapt the previous
picture to denote the effective Hamiltonian:\\
\setlength{\unitlength}{0.8mm}
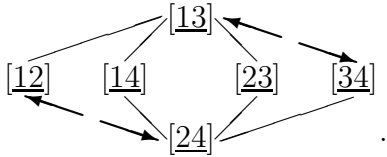
\begin{picture}(72,35)(-10,0)
\thinlines
\put(16,18){\line(3,1){22.14}}  % [12] to [13]
\put(32,18){\line(1,1){7}}  % [14] to [13]
\put(54,18){\line(-1,1){7}}  % [23] to [13]
\put(32,12){\line(1,-1){7}}  % [14] to [24]
\put(54,12){\line(-1,-1){7}}  % [23] to [24]
\put(70,12){\line(-3,-1){22.14}}  % [34] to [24]
\thicklines
\put(61,21){\vector(3,-1){9.5}} % [34] to [13]
\put(58,22){\vector(-3,1){9.5}} % [34] to [13]
\put(25,9){\vector(-3,1){9.5}} % [12] to [24]
\put(28,8){\vector(3,-1){9.5}} % [12] to [24]
\put(16,15){\makebox(0,0){[\underline{12}]}}
\put(32,15){\makebox(0,0){[\underline{14}]}}
\put(54,15){\makebox(0,0){[\underline{23}]}}
\put(70,15){\makebox(0,0){[\underline{34}]}}
\put(43,25){\makebox(0,0){[\underline{13}]}}
\put(43,5){\makebox(0,0){[\underline{24}]}}
\put(75,5){\makebox(0,0){.}}
\end{picture}\\
Here the thin lines are still matrix elements $-t$. The thick lines
have been cut [we do not allow these hops in the effective Hamiltonian] and
replaced by arrows, indicating diagonal matrix elements, which
in this simple case all become $+t$, because we have chosen equal
weights for all the states in the trial wave function.
The [nondegenerate] ground state of this effective Hamiltonian is
\begin{eqnarray} |\psi_0^{\mbox{\scriptsize eff}}\rangle & = &
|0.165([\underline{12}] + [\underline{34}]) + \nonumber \\
 & + & 0.448([\underline{13}] + \underline{24}]) + 0.523([\underline{14}] +
[\underline{23}])\rangle \end{eqnarray}
with energy -1.709t. Note that, e.g., the states $[\underline{12}]$ and
$[\underline{24}]$ do {\em not\/} have the same wave function in this ground
state, while they do in the trial state. As one could have expected from
symmetry considerations, the wave function is the same in states that have an
equivalent position in the picture, i.e. occur symmetrically in the effective
Hamiltonian.
States that are connected via the boundary do not in general have such
symmetry, and thus there is no reason to expect that they would
obey~(\ref{leverrule}). Note also that the energy of the effective ground state
is above the ground-state energy of the true problem,
as it should be according to our proof that it is an upper bound for that
energy.

\ifpreprintsty
\begin{table}
\begin{tabular}{|c|c|c|r@{}l|r@{}l|r@{}l|}
\hline
 system & $U$ & trial & \multicolumn{6}{c|}{energies} \\
 \cline{4-9}
 & & type & \multicolumn{2}{c|}{MF} & \multicolumn{2}{c|}{FN} &
\multicolumn{2}{c|}{exact} \\
 \hline \hline
 square & $0$ & & -4 & & -4 & & -4 & \\
 2$\uparrow$~~2$\downarrow$ & $1$ & AF & -3 & .2855 & -3 & .3172 & -3 & .3409
\\
 \hline
 cube & $0$ & & -9 & & -9 & & -9 & \\
 4$\uparrow$~~1$\downarrow$ & $1$ & H & -8 & .5 & -8 & .5419 & -8 & .5420 \\
  & $6$ & H & -6 & & -7 & .2508 & -7 & .2533 \\
  & $6$ &  & -6 & .0701 & -7 & .2424 & -7 & .2533 \\
  & $10$ & H & -4 & & -6 & .8400 & -6 & .8442 \\
  & $10$ & AF & -4 & .2551 & -6 & .7476 & -6 & .8442 \\
  & $10$ &  & -5 & .3271 & -6 & .7637 & -6 & .8442 \\
 \hline
 cube & $0$ & & ~-12 & & ~-12 & & ~-12 & \\
 4$\uparrow$~~4$\downarrow$ & $1$ & H & -10 & & -10 & .1148~ & -10 & .1188~ \\
  & $~2.5~$ & H & -7 & & -7 & .7257 & -7 & .7510 \\
  & $2.5$ & AF & -7 & .0061~ & -7 & .6942 & -7 & .7510 \\
  & $10$ & H & 8 & & -2 & .6597 & -2 & .8652 \\
  & $10$ & AF & -2 & .3113 & -2 & .6382 & -2 & .8652 \\
 \hline
\end{tabular}
{}~\\
\caption{\label{tab-results} Comparison of the energies obtained for three
different systems by means of self-consistent mean-field (MF), fixed-node (FN),
and exact calculations\protect\cite{exact}. All values are given in units of
$t$. }
\end{table}
\else
\end{multicols}
\fi

\begin{thebibliography}{99}
\bibitem{statacc} For a review and recent references, see e.g. W. von der
Linden, Phys. Rep. {\bf 220}, 53 (1992); or H. de Raedt and W. von der Linden,
in {\em The Monte Carlo Method In Condensed Matter Physics\/}, edited by K.
Binder (Springer Verlag, Berlin, 1992).
\bibitem{bemmel} H.J.M. van Bemmel, D.F.B. ten Haaf, W. van Saarloos,
J.M.J. van Leeuwen, and G. An, Phys. Rev. Lett. {\bf 72}, 2442 (1994).
\bibitem{ceperley} D.M. Ceperley and B.J. Alder, Phys. Rev. Lett. {\bf 45}, 566
(1980); and Science {\bf 231}, 555 (1986).
\bibitem{fixedphase} G. Ortiz, D.M. Ceperley and R.M. Martin, Phys. Rev. Lett.
{\bf 71}, 2777 (1993).
\bibitem{tricep} N. Trivedi and D.M. Ceperley, Phys. Rev. B {\bf 41}, 4552
(1990).
\bibitem{anderson} J.B. Anderson, J. Chem. Phys. {\bf 63}, 1499 (1975); {\bf
65}, 4122 (1976).
\bibitem{an} G. An and J.M.J. van Leeuwen, Phys. Rev. B {\bf 44}, 9410 (1991).
\bibitem{mitas} L. Mitas, E.L. Shirley, and D.M. Ceperley, J. Chem. Phys. {\bf
95}, 3467 (1991).
\bibitem{exact} The FN and exact results have been obtained by means of
numerical routines to exactly diagonalize a matrix of dimension $36\times 36$
(for the square with 2$\uparrow$ and 2$\downarrow$), $560\times 560$ (cube with
4$\uparrow$, 1$\downarrow$), and $4900\times 4900$ (cube with 4$\uparrow$,
4$\downarrow$), respectively. In the latter case, group theory was used to
reduce the matrix to blocks of $1896\times 1896$ or smaller.
\bibitem{martin} R. Martin, {\em Energy functionals for electronic structure
calculations\/} (preprint).

\end{thebibliography}
\end{document}